\documentclass[twocolumn]{aastex701}
\usepackage{amsmath}
\usepackage{booktabs}
\usepackage{colortbl}
\definecolor{tabblue}{RGB}{31,119,180}
\graphicspath{{}}

\shorttitle{JEPA for astronomy}
\shortauthors{}

\begin{document}

\title{JEPA for astronomy: \\ multi-survey spectral alignment puts stellar parameters on the leading principal axes}

\correspondingauthor{Xiaosheng Zhao}
\email{xzhao113@jhu.edu}

\author[0000-0002-8328-1447]{Xiaosheng Zhao}
\affiliation{William H.~Miller III Department of Physics \& Astronomy, \\ The Johns Hopkins University, Baltimore, MD 21218, USA}
\email{xzhao113@jhu.edu}

\author[0000-0002-4013-1799]{Rosemary F.~G. Wyse}
\affiliation{William H.~Miller III Department of Physics \& Astronomy, \\ The Johns Hopkins University, Baltimore, MD 21218, USA}
\email{wyse@jhu.edu}

\author[0000-0001-5082-9536]{Yuan-Sen Ting}
\affiliation{Department of Astronomy, The Ohio State University, 140 West 18th Avenue, Columbus, OH 43210, USA}
\affiliation{Center for Cosmology and AstroParticle Physics (CCAPP), The Ohio State University, Columbus, OH 43210, USA}
\affiliation{Max-Planck-Institut f\"ur Astronomie, K\"onigstuhl 17, D-69117 Heidelberg, Germany}
\affiliation{Asia School of Business, No. 1 \& 11, Jalan Dato Onn, 50480 Kuala Lumpur, Malaysia}
\email{ting.74@osu.edu}

\author[0000-0002-4108-3282]{Alexander S. Szalay}
\affiliation{William H.~Miller III Department of Physics \& Astronomy, \\ The Johns Hopkins University, Baltimore, MD 21218, USA}
\affiliation{Department of Computer Science, The Johns Hopkins University, Baltimore, MD 21218, USA}
\email{szalay@jhu.edu}

\begin{abstract} Spectra from different surveys carry different systematics and are not calibrated against one another. Multi-survey foundation models for stellar spectra are then built by aligning embeddings of the same star observed across surveys. Most in astronomy follow the recipes established for language and vision, aligning with a contrastive loss that separates every star from the others. However, this does not ensure that the stellar physics is encoded in a form that a simple read-out can recover. Holding the cross-survey architecture fixed and varying only the alignment loss, we measure what an objective encodes from the principal-component geometry of the Gaia BP/RP (XP) spectra and from a linear probe of the stellar parameters. In a properly aligned embedding, the leading two principal components form a plane that recovers an analog of the Kiel diagram, the third component tracks [Fe/H], and [$\alpha$/Fe] varies coherently across the plane, such that the chemically defined Galactic disks and accreted populations occupy distinct loci. We show that legibility, the ease of reading the information out, is highest for a training objective that never pushes stars apart, and instead constrains only the overall distribution of each survey's embeddings (JEPA). JEPA also exceeds the contrastive alternative at the same architecture, and the performance improves further as more surveys (Gaia\,XP$+$LAMOST\,LRS$+$DESI$+$APOGEE) are included. Predicted abundances for 8.0 million XP giants reproduce the main [$\alpha$/Fe]–[Fe/H] structure of the Milky Way. These results indicate that the choice of alignment objective, not alignment alone, determines whether stellar labels are linearly recoverable.
\end{abstract}


\section{Introduction}
\label{sec:intro} Repeated observation of the same object underlies a broad class of astronomical inference. Parallax recovers, from lines of sight taken at different points around the Sun, a distance that no single sight contains on its own. Multi-survey spectroscopy poses the same problem in a new form. One star observed by Gaia XP \citep{2023A&A...674A...2D,2023A&A...674A...1G}, LAMOST \citep{2012RAA....12.1197C,2012RAA....12..723Z}, DESI \citep{2016arXiv161100036D}, and APOGEE \citep{2017AJ....154...94M} yields four very different spectra, differing in resolution, wavelength coverage, and other instrumental signatures, while all four arise from the same single star with one value of the temperature, one surface gravity, and one chemical composition. The insights provided by the spectra must agree on the stellar physical parameters and differ only through the constraints imposed by the instrumental configurations. How best to recover that underlying physical state from the agreement between the surveys alone, with few or no stellar labels, is the problem addressed here.

A foundation model \citep{2021arXiv210807258B,2026arXiv260802573Z} addresses this through its \emph{internal representation}. Trained once on broad data, it maps each spectrum to a fixed-length vector, or \emph{embedding}, in which the representation is expressed. Later measurements can then be made from the embedding at low cost, recovering much of the accuracy a purpose-built supervised model would need abundant reference labels to reach. The advantage is largest in the label-scarce regime typical of astrophysics, where reliable stellar parameters are available only for the minority of stars observed at higher spectral resolution. These abilities are often attributed to the transferability of the foundation model. Transferability is conventionally quantified by the cost of fitting a small predictor on the fixed embedding, a measure that usually conflates whether the information is present with how simply it can be extracted. 

However, where the information sits matters as much as whether it is there, since the directions of the greatest variance in an embedding are not necessarily those that carry the parameter of interest \citep{2026arXiv260802573Z}. The most useful case is one in which the physical factors are laid out simply enough that a linear read-out suffices, so that a star's position in the embedding, or a linear transformation of it, is itself a statement about its parameters. We call this property legibility: the ease with which information can be read out of an embedding, here by a linear map. The distinction can be the familiar one between a Kiel diagram, where position on the diagram carries effective temperature and surface gravity, and a list of fluxes from which the same quantities can be obtained only by fitting. An embedding of the second kind may hold the physics well, but each measurement has to be learned from it rather than read off it.

Building such a representation across several surveys is harder, but the gain is also larger. Aligning several such views contrastively tends to produce embeddings in which each survey clusters in its own region of the shared space rather than interleaving with the others (known in machine learning as the ``modality gap'', \citealp{liang2022modalitygap}). Where alignment succeeds, however, each survey retains and reorganizes the information it shares with the others, and performance could improve even for stars observed by only one survey, as seen in the information sharing between text and vision \citep{yuan2021multimodal}. Cross-modal and omnimodal foundation models in astronomy already align images, spectra, light curves, and metadata across surveys \citep{Parker2024,Parker2025,Rizhko2024}. For stellar spectra, \citet{Zhao2025} and \citet{Buck2024} align low-resolution Gaia XP with higher-resolution surveys. Most of them align contrastively, following CLIP \footnote[1]{We use CLIP throughout for the contrastive objective, whose loss is InfoNCE \citep{Oord2018,Radford2021}.}, so that the alignment is carried by a loss requiring negative examples. Non-contrastive joint-embedding objective, such as those used in JEPA \citep{lecun2022path,Assran2023}, have been much less explored in multi-survey astronomy.

We test whether such an aligned embedding organizes itself along the known axes of stellar physics and Galactic chemical evolution, including the thin and thick disks, the accreted debris, and the very metal-poor tail. We read that geometry directly from the embedding's leading variance and quantify it with a linear probe, while the representation itself is trained without labels. Holding the architecture fixed, we vary the alignment loss and the number of surveys aligned, and ask which ingredients make this structure more linearly accessible.

\section{Methods} \label{sec:methods}

Cross-survey models are usually assembled in two stages, following the construction that large-language and vision models adopted. Each survey is first encoded, most often by an autoencoder (AE) trained to reconstruct its own spectra, and the resulting representations are then tied together across surveys by a contrastive alignment loss. Neither stage ensures that the physics is left in a form that can be read easily. A decoder in the first stage is graded on every pixel it is asked to reproduce, and in a stellar spectrum a large part of the pixel-to-pixel variance is not due to the properties of the star at all, but line-spread function, flux calibration, fiber and sky residuals, and noise. A model required to reconstruct the input spectrum must therefore retain all of it, and the directions of largest variance in such an embedding describe the instrument as much as the star.  

\subsection{What an alignment objective must do}

Alignment is then designed at the second stage after encoding to map the information to a more interpretable and physical space shared across surveys, and it is carried mainly by the training objective. Such an objective has two tasks: 1) to pull matched views of the same star together, and 2) to prevent the model from meeting that demand in a trivial way. An objective that does only the first admits a degenerate solution in which the encoder maps every spectrum to the same vector and discards all spectral information, a failure mode called \emph{collapse}. The objectives compared here differ mainly in how they conduct the second task.

Contrastive training is the standard solution, used by CLIP and its astronomical variants \citep{Oord2018,Radford2021,Parker2024,Zhao2025}. The loss function prevents collapse by also pushing the embeddings of different stars apart. Every other star in the training batch acts as a \emph{negative} to avoid a trivial constant solution. This remedy is somewhat artificial in that \emph{collapse} is a defect of the objective rather than a property of the stars themselves, and separating stars is an operation introduced to prevent the \emph{collapse}, not a requirement imposed by the underlying physics. It is also poorly suited to spectroscopy in particular because large surveys contain many stars that are very similar in effective temperature, surface gravity, and composition, so this operation could drive a solar twin away from the Sun for no reason other than the loss function. Which stars serve as negatives depends on how the data happen to be arranged in the training batches, so part of the negative-penalty a star receives is an accident of datasets shuffling, and training is limited by the batch size for the same reason \citep{Oord2018}.

\begin{figure}
\centering
\includegraphics[width=\columnwidth]{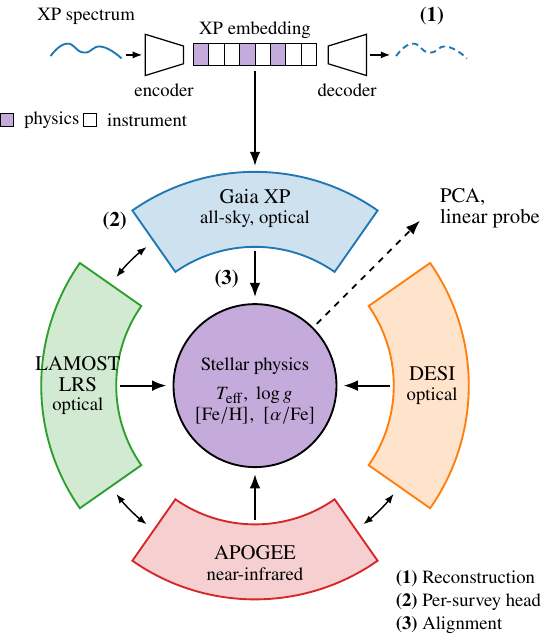}
\caption{Multi-survey alignment. Embeddings from reconstructive autoencoders are mapped by a survey-specific head into a shared embedding space. Alignment is performed in a subsequent projection space (not shown for simplicity), which is discarded after training. Since the views share the same star, the common information is dominated by the underlying physical state (center), while instrument-specific information is suppressed. PCA and linear probes are then applied to aligned embeddings from a specific survey (e.g., Gaia). \label{fig:arch}}
\end{figure}

\subsection{The JEPA objective}
\label{sec:objective}

In a different approach to the second task, \emph{collapse} can be ruled out without pushing stars apart at all. Some methods rely on architectural asymmetry \citep{grill2020byol,caron2021dino,oquab2023dinov2}, whereas others constrain the statistics of the embedding taken as a whole (Barlow Twins, \citealp{2021arXiv210303230Z}; VICReg, \citealp{Bardes2021}), so that the trivial solution is excluded by population-level statistics rather than pairwise separation. The latter route leads to the JEPA (Joint-Embedding Predictive Architecture; \citealp{lecun2022path,Assran2023}) model family, whose \emph{distribution-matching} objectives prevent \emph{collapse} with no negatives at all. We use its rectified-$L_p$ member which applies the RDMReg objective \citep{Kuang2026} and refer to it as JEPA throughout this paper. The variants we tried, including the Gaussian-target member SIGReg \citep{Balestriero2025}, and a similar objective VICReg \citep{Bardes2021}, are described in Appendix~\ref{sec:losses}.

The JEPA objective acts on a projected embedding $\mathbf{z}=g(\mathbf{h}) \in \mathbb{R}^{d}$ ($d=128$), where $\mathbf{h}$ is the 768-dimensional output of the per-survey head and $g$ is a projector used only during training. The first task, pulling the matched views together, is an invariance term
\begin{equation}
\mathcal{L}_{\rm inv}=\bigl\langle d^{-1}\lVert\mathbf{z}_a-\mathbf{z}_b\rVert^2\bigr\rangle_{\rm pairs},
\label{eq:inv}
\end{equation}
the mean squared difference per coordinate between the two views of the same star, averaged over stars and over the survey pairs (with the weights given in Appendix~\ref{sec:app}). The second task is carried by a regularizer $\mathcal{R}$, a term that penalizes the degenerate solution, added to give $\mathcal{L}=\mathcal{L}_{\rm inv}+\lambda\mathcal{R}$, in which the ratio $\lambda$ is tunable to balance loss terms.

Distribution matching makes $\mathcal{R}$ a comparison between the projected embeddings and a specific target distribution $T$. Comparing distributions directly in all $d$ dimensions is impractical from a finite batch of stars, so the comparison is \emph{sliced}, meaning that it is carried out along uniformly random one-dimensional projections,

\begin{equation}
\mathcal{R}=\mathbb{E}_{\mathbf{u}\sim\mathrm{Unif}(\mathbb{S}^{d-1})}\bigl[\,D\bigl(\mathbb{P}_{\mathbf{u}^{\!\top}\mathbf{z}},\;\mathbb{P}_{\mathbf{u}^{\!\top}\mathbf{t}}\bigr)\bigr],
\label{eq:sliced}
\end{equation}
where $\mathbf{u}$ is a random unit vector, $\mathbf{u}^{\!\top}\mathbf{z}$ is the single number obtained by projecting each star's embedding onto it, and $\mathbf{t}$ is sampled from $T$. Thus, instead of asking whether two complicated $d$-dimensional clouds have the same distribution, the loss repeatedly asks whether they look the same when viewed from random directions. JEPA takes $D$ to be the 2-Wasserstein distance, evaluated by sorting two sets of projected values and comparing them pairwise. Its target $T$ is a product of rectified generalized Gaussian distributions obtained by drawing generalized-Gaussian variables with density $\propto \exp(-|x-\mu|^p/\sigma^p)$, with $\sigma$ set so that each variable has unit variance, and mapping negative values to zero. For $p=1$, the underlying continuous distribution is Laplace.

The shape of that target has a physical reading. It asks that, in each embedding coordinate, many stars are pushed to zero while only a fraction stand out with positive values, so each coordinate behaves like the strength of a feature that is present in some stars and absent in others.

Equation~\ref{eq:sliced} rules out the possibility of \emph{collapse} occurring, since a constant embedding would make $\mathbf{u}^{\!\top}\mathbf{z}$ a point mass in every direction, which cannot match a target distribution of fixed and nonzero width.

\subsection{Models and probes}

Figure~\ref{fig:arch} shows the setup, in which the surveys provide independent views of the same stars. Each survey is encoded by a frozen backbone, a pre-trained encoder held fixed throughout, followed by a trainable head and a projector used only during training. Every objective uses identical backbones, and only the alignment loss differs, so that the comparison changes the pre-training alone while holding the evaluation process fixed.

We train each objective in a two-survey and four-survey configuration. The JEPA models use a Laplace ($p=1$) distribution-matching target \citep{Kuang2026}. The contrastive control (CLIP) carries a learned temperature $\tau$ \citep{Oord2018,Radford2021}. The unaligned autoencoder has no cross-survey term and isolates what reconstruction alone provides, serving as the no-alignment floor. The tables label these three JEPA, CLIP, and AE. Detailed training setup is described in Appendix~\ref{sec:app} and Figure~\ref{fig:arch_details}.

The embedding dimension is fixed at 768 and is not tuned, and we investigate each embedding in two ways. First, we compute a principal-component analysis (PCA; \citealp{hotelling1933analysis}) of the embedding and inspect its leading plane. Second, we train two supervised probes at a range of labeled-sample sizes and evaluate them on the same withheld stars for every objective: a linear ridge regressor \citep{10.1002/wics.14}, which is a linear regression with an $\ell_2$ penalty on the coefficients, and a nonlinear multilayer perceptron (MLP), a small full-connected network. Both are configured as in Appendix~\ref{sec:app}. Before probing, we standardize each of the 768 embedding coordinates using means and variances from the training split alone, which are then applied unchanged to the withheld test set. PCA, the ridge probe and the classifiers of Section~\ref{sec:coord} use \textit{scikit-learn} \citep{JMLR:v12:pedregosa11a}.

The performance difference between the two probes is diagnostic because, if  a nonlinear probe can recover a parameter but a linear one cannot, that parameter is present in the embedding but entangled across many embedding coordinates. Scores are quoted as the coefficient of determination $R^2$, computed after 3$\sigma$ clipping so that a few badly-fit stars cannot dominate the comparison.

\section{Data} \label{sec:data}

\begin{table}[!t]
\centering
\caption{Cross-matched observations used for alignment, and the labeled sample used to evaluate the Gaia XP embeddings. \label{tab:data}}
\begin{tabular}{@{}llrl@{}}
\toprule
\multicolumn{2}{@{}l}{Survey pair} & Matched pairs & Used in \\
\midrule
Gaia XP & LAMOST LRS & $820{,}568$ & all models \\
Gaia XP & APOGEE    & $454{,}355$ & four-survey \\
Gaia XP & DESI      & $313{,}212$ & four-survey \\
DESI    & LAMOST LRS & $260{,}063$ & four-survey \\
APOGEE  & LAMOST LRS & $145{,}195$ & four-survey \\
APOGEE  & DESI       & $16{,}732$  & four-survey \\
\midrule
\multicolumn{2}{@{}l}{Labeled sample} & $113{,}218$ & \\
\multicolumn{2}{@{}l}{\quad Training}  & $101{,}891$ & \\
\multicolumn{2}{@{}l}{\quad Withheld}  & $11{,}327$  & \\
\bottomrule
\end{tabular}
\end{table}

Alignment is trained on cross-matched observations of the same star, listed in Table~\ref{tab:data}. The two-survey models use the pairs of Gaia XP DR3 \citep{2023A&A...674A...2D} and LAMOST LRS DR11 \citep{2026yCat.5162....0L}, and the four-survey model adds DESI DR1 \citep{2026AJ....171..285D} and APOGEE DR17 \citep{2022ApJS..259...35A} and trains on all six pairings. Each XP spectrum is $343$ flux points divided by the flux at $550$\,nm. All embeddings analyzed here are those of the Gaia XP model head, evaluated on a common set of stars with reference labels.

For interpreting the embedding geometry we assign each star to at most one of five chemically defined categories, using reference abundances that come mostly from APOGEE, supplemented for [Fe/H] by metal-poor compilations \citep{2019MNRAS.484.2166S, 2022ApJ...931..147L, 2024ApJ...974..192H}. The categories are \emph{thin disk} ($\mathrm{[Fe/H]}\in[-0.7,0.5]$, $\mathrm{[\alpha/Fe]}\in[-0.02,0.12]$; $46{,}627$ stars); \emph{thick disk} ($\mathrm{[Fe/H]}\in[-0.9,-0.1]$, $\mathrm{[\alpha/Fe]}\in[0.18,0.35]$; $19{,}340$); \emph{accreted-star candidates} (predominantly, but not exclusively, likely from Gaia--Enceladus/Sausage), selected chemically in the [Al/Fe]--[Mg/Mn] plane using the criterion of \citet{Horta2022}, namely $\mathrm{[Mg/Mn]}\ge0.25$ for $\mathrm{[Al/Fe]}<-0.07$, and $\mathrm{[Mg/Mn]}\ge4.25\,\mathrm{[Al/Fe]}+0.5475$ ($9{,}292$ stars); \emph{very metal-poor} (VMP, $\mathrm{[Fe/H]}{<}-2$ with a valid [$\alpha$/Fe]; $143$); and \emph{low-$\alpha$\,\&\,metal-poor} ($\mathrm{[Fe/H]}\in[-1.5,-0.5]$, $\mathrm{[\alpha/Fe]}{<}0.05$; $5{,}470$). The categories overlap, and each star therefore takes the first matching category in the following order: very metal-poor, accreted, thick disk, low-$\alpha$ and metal-poor, then thin disk. Of the $113{,}218$ labeled stars (training and withheld sets combined; Table~\ref{tab:data}), $80{,}872$ are classified and $29\%$ remain unclassified.

\begin{figure*}
\centering
\includegraphics[width=0.86\textwidth]{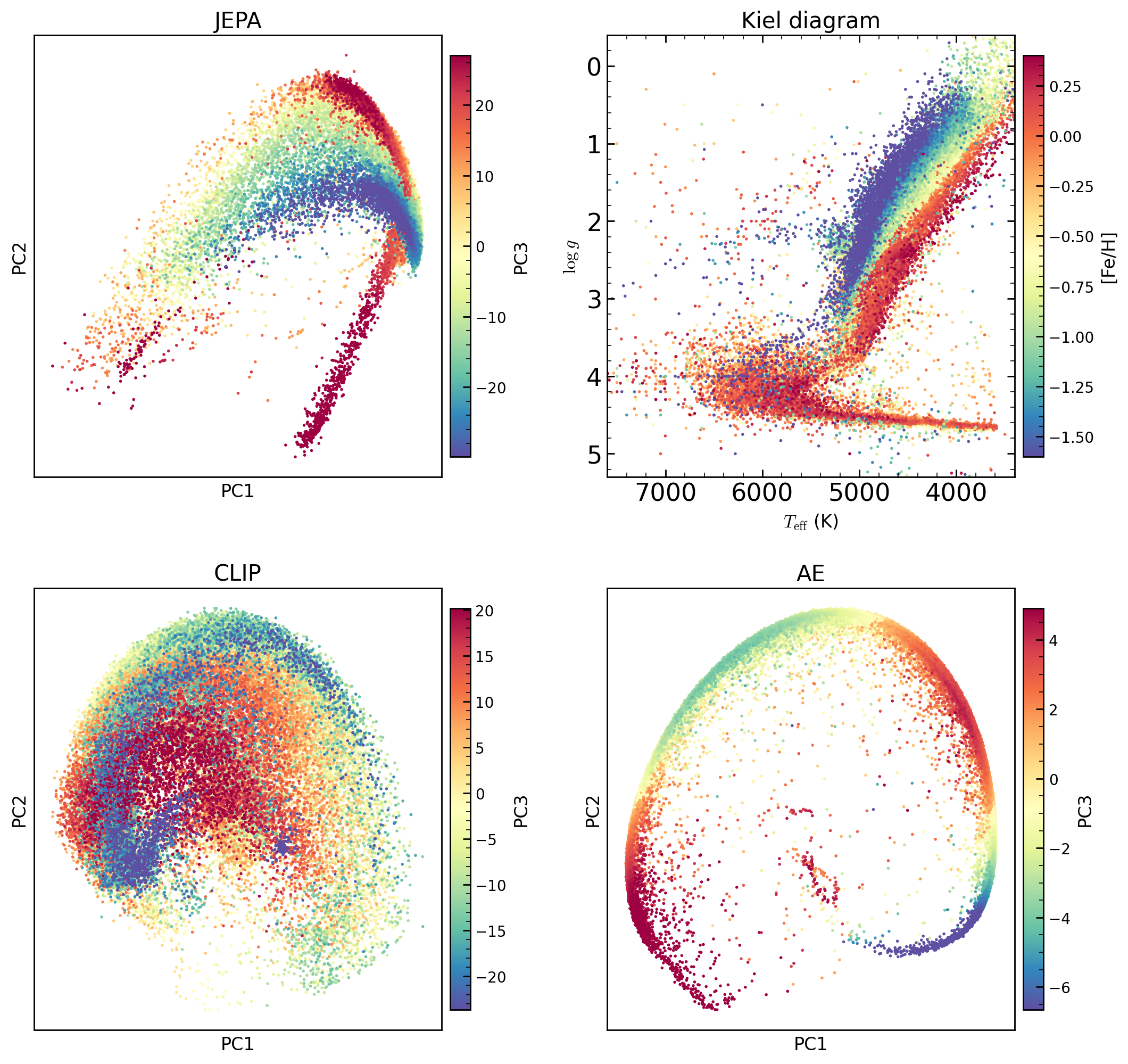}
\caption{Linear legibility of [Fe/H]. Each panel shows the first two principal components, PC1 and PC2, colored by PC3, from embeddings of two alignment models (JEPA, top left; CLIP, bottom left) and one unaligned autoencoder (AE, bottom right). The APOGEE Kiel diagram colored by [Fe/H] (top right) is shown as the physical anchor. Components are computed independently for each panel and are in arbitrary units. The same $40{,}000$ stars are shown throughout. Reconstruction alone (AE) leaves [Fe/H] unordered. In contrast, alignment recovers the ordering, and the objective determines how sharply it is recovered (JEPA vs CLIP). \label{fig:obj}}
\end{figure*}

\begin{table*}
\centering
\caption{Recovery of stellar parameters from each 768-d embedding, at two labeled-sample sizes. \label{tab:r2}}
\begin{tabular}{llccccc}
\toprule
parameter & $N$ & JEPA (4 survey) & CLIP (4 survey) & JEPA (2 survey) & CLIP (2 survey) & AE \\
 & & \footnotesize ridge\,/\,MLP & \footnotesize ridge\,/\,MLP & \footnotesize ridge\,/\,MLP & \footnotesize ridge\,/\,MLP & \footnotesize ridge\,/\,MLP \\
\midrule
$T_{\rm eff}$ & 200 & \cellcolor{tabblue!16}0.93\,/\,0.97 & \cellcolor{tabblue!8}0.86\,/\,0.96 & \cellcolor{tabblue!28}\textbf{0.94}\,/\,0.96 & 0.68\,/\,0.97 & 0.74\,/\,0.93 \\
 & 10{,}000 & \cellcolor{tabblue!16}0.95\,/\,0.98 & \cellcolor{tabblue!8}0.93\,/\,0.98 & \cellcolor{tabblue!28}\textbf{0.96}\,/\,0.99 & \cellcolor{tabblue!8}0.93\,/\,0.98 & \cellcolor{tabblue!16}0.95\,/\,0.98 \\
\midrule
$\log g$ & 200 & \cellcolor{tabblue!28}\textbf{0.95}\,/\,0.96 & \cellcolor{tabblue!16}0.93\,/\,0.95 & \cellcolor{tabblue!16}0.93\,/\,0.95 & \cellcolor{tabblue!8}0.88\,/\,0.94 & 0.70\,/\,0.90 \\
 & 10{,}000 & \cellcolor{tabblue!28}\textbf{0.98}\,/\,0.98 & \cellcolor{tabblue!16}0.97\,/\,0.98 & \cellcolor{tabblue!16}0.97\,/\,0.98 & \cellcolor{tabblue!8}0.96\,/\,0.98 & 0.94\,/\,0.96 \\
\midrule
{[Fe/H]} & 200 & \cellcolor{tabblue!28}\textbf{0.90}\,/\,0.91 & \cellcolor{tabblue!16}0.87\,/\,0.90 & \cellcolor{tabblue!8}0.86\,/\,0.90 & 0.82\,/\,0.88 & 0.69\,/\,0.75 \\
 & 10{,}000 & \cellcolor{tabblue!28}\textbf{0.95}\,/\,0.96 & \cellcolor{tabblue!16}0.94\,/\,0.95 & \cellcolor{tabblue!8}0.93\,/\,0.95 & 0.92\,/\,0.95 & 0.87\,/\,0.92 \\
\midrule
{[$\alpha$/Fe]} & 200 & \cellcolor{tabblue!28}\textbf{0.61}\,/\,0.70 & \cellcolor{tabblue!16}0.60\,/\,0.67 & \cellcolor{tabblue!8}0.55\,/\,0.61 & 0.53\,/\,0.62 & 0.31\,/\,0.29 \\
 & 10{,}000 & \cellcolor{tabblue!28}\textbf{0.79}\,/\,0.84 & \cellcolor{tabblue!16}0.75\,/\,0.82 & \cellcolor{tabblue!8}0.74\,/\,0.82 & 0.71\,/\,0.81 & 0.58\,/\,0.74 \\
\bottomrule
\end{tabular}
\tablecomments{Each cell is \emph{linear ridge\,/\, nonlinear probe}, as $3\sigma$-clipped $R^2$ on the withheld stars, shaded per row with darkest and bold marking the best ridge value.}
\end{table*}

\section{The alignment objective sets the linear legibility of chemistry} \label{sec:linear}

The embeddings from the following three models compared here differ only in the loss used to train them. The unaligned autoencoder carries no cross-survey term and isolates what reconstruction alone provides, the contrastive model adds alignment with negative examples, and the JEPA model adds alignment without them.

Figure~\ref{fig:obj} colors the plane of the first two principal components by the third, for two objectives at the four-survey scale and for the unaligned autoencoder. The three geometries differ. For the four-survey JEPA model (top left) the leading plane of the first two principal components is a Kiel-shaped image of the true Kiel diagram, and the third component is an ordered gradient closely tracking [Fe/H]. For the four-survey CLIP (bottom left) the same axis is only partially ordered, as the iron-abundance gradient is present but diluted. For the unaligned autoencoder (bottom right) the leading three components show no such monotonic ordering. Alignment therefore places chemistry on directions that a linear probe can read, and how far it does so depends on the objective.

An ordered-looking gradient need not carry much of a parameter, so the geometry must be checked against what a particular choice of probe recovers. Table~\ref{tab:r2} reports the coefficients of determination ($R^2$) of linear ridge alongside the nonlinear MLP, for a label-scarce case at $N=200$ and a well-sampled one at $N=10{,}000$. A bootstrap over the withheld stars gives a per-cell ridge uncertainty of 0.001--0.011, largest for [$\alpha$/Fe] at smaller $N$, so the differences between objectives are well-resolved. 

At labeled samples $N{=}10{,}000$ the ridge recovery of [$\alpha$/Fe] falls from $R^2{=}0.79$ for the four-survey JEPA model to $0.58$ for the unaligned autoencoder, with the CLIP and two-survey models in between. The difference in performance measured by $\Delta R^2$ between the linear and nonlinear probes widens from $0.05$ for the best-aligned model to $0.16$ for the unaligned one. The advantage of the four-survey JEPA model over the unaligned autoencoder grows where labels are scarcest, with the ridge performance difference reaching $0.30$ in [$\alpha$/Fe] and $0.21$ in [Fe/H] at $N{=}200$.

Those scores use all 768 dimensions, whereas the figure shows only the leading plane, so restricting the linear fit to the leading components tests whether the visible ordering is where the linearly accessible content sits. A ridge fit on the first three components alone explains $R^2{=}0.69$ of [Fe/H] for the four-survey JEPA but only $0.02$ for the unaligned autoencoder, a separation far wider than in the full embedding. The CLIP model reaches $0.68$, so its leading components do carry the iron abundance, but spread over a nearly flat eigenvalue spectrum whose flatness of the top ten eigenvalues, measured as geometric over arithmetic mean, is 0.83, vs 0.60 for the JEPA model. The single-component [Fe/H] axis of the CLIP model is less well defined, with the third component alone reaching $R^2=0.58$, against $0.68$ for the JEPA model. When restricted to the leading two-component plane, the same fit yields $R^2\le0.10$, indicating that the iron abundance sits linearly just off the Kiel plane rather than in it. 

The number of surveys aligned improves the score while maintaining the validity of the comparison. The four-survey JEPA model outperforms its two-survey counterpart, with higher linear $R^2$ for [Fe/H] and [$\alpha$/Fe] across $N$ (Table~\ref{tab:r2}) and better-ordered iron-abundance structure in the leading principal components. The gain from two to four surveys does not arise from additional information for the same XP wavelength coverage, since APOGEE observes the near-infrared $H$-band ($\approx1.5$--$1.7\,\mu$m, outside XP's $0.336$--$1.021\,\mu$m range), while the LAMOST/DESI partners are higher-resolution optical, both used here through a blue-arm window (0.4--0.56\,$\mu$m). The gain is therefore complementary. Alignment lets the higher-resolution partners locate the chemistry within XP's own low-resolution spectra and place it on linearly readable axes.

Gaia XP spectra themselves therefore carry iron abundance and $\alpha$ information \citep[e.g.][]{2024ApJS..272....2L,Laroche2024}, and alignment supplies not the information itself but its placement along directions that a linear read-out can follow. Different objectives vary in how readily they make that information linearly accessible, adding surveys sharpens it further. 

\begin{figure*}
\centering
\includegraphics[width=0.98\textwidth]{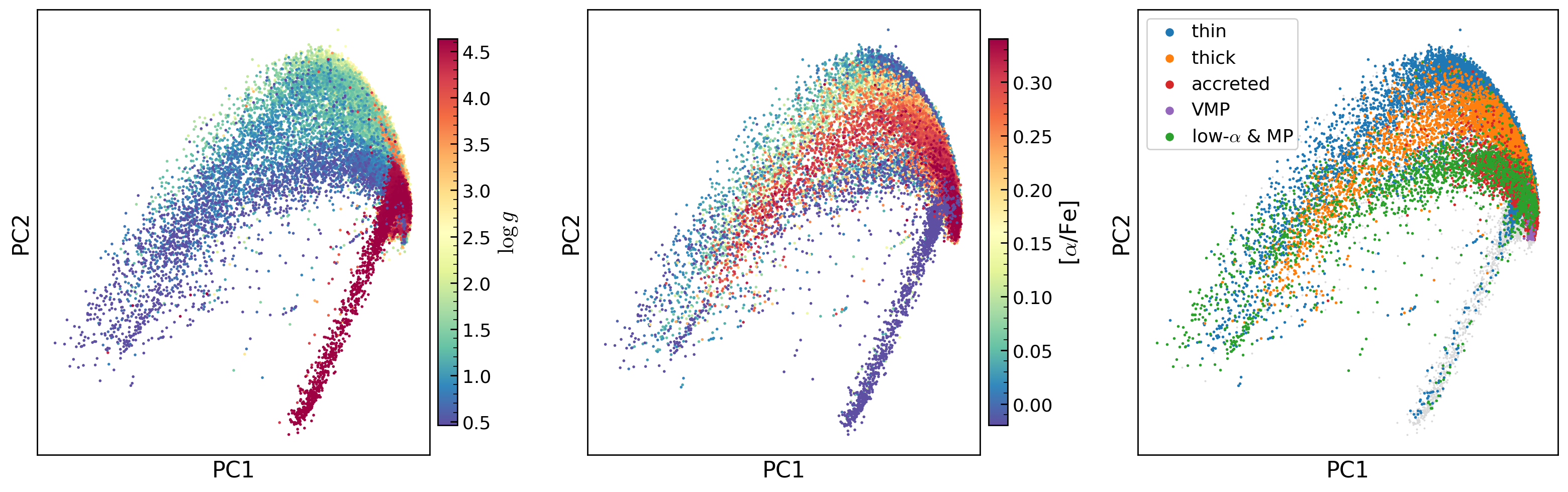}
\caption{Gravity and chemistry encoding. The same PC1--PC2 plane as in Figure~\ref{fig:obj} is shown for the JEPA model, colored by surface gravity $\log g$ (left), by [$\alpha$/Fe] (middle), and by chemical category (right, where gray marks unclassified stars and thin/thick/accreted/VMP/low-$\alpha$\,\&\,MP contain $46{,}627/19{,}340/9{,}292/143/5{,}470$ stars). The plane is ordered by surface gravity. The chemical populations occupy distinct bands whose [$\alpha$/Fe] matches their chemical definition. \label{fig:coord}}
\end{figure*}

\begin{figure*}
\centering
\includegraphics[width=\textwidth]{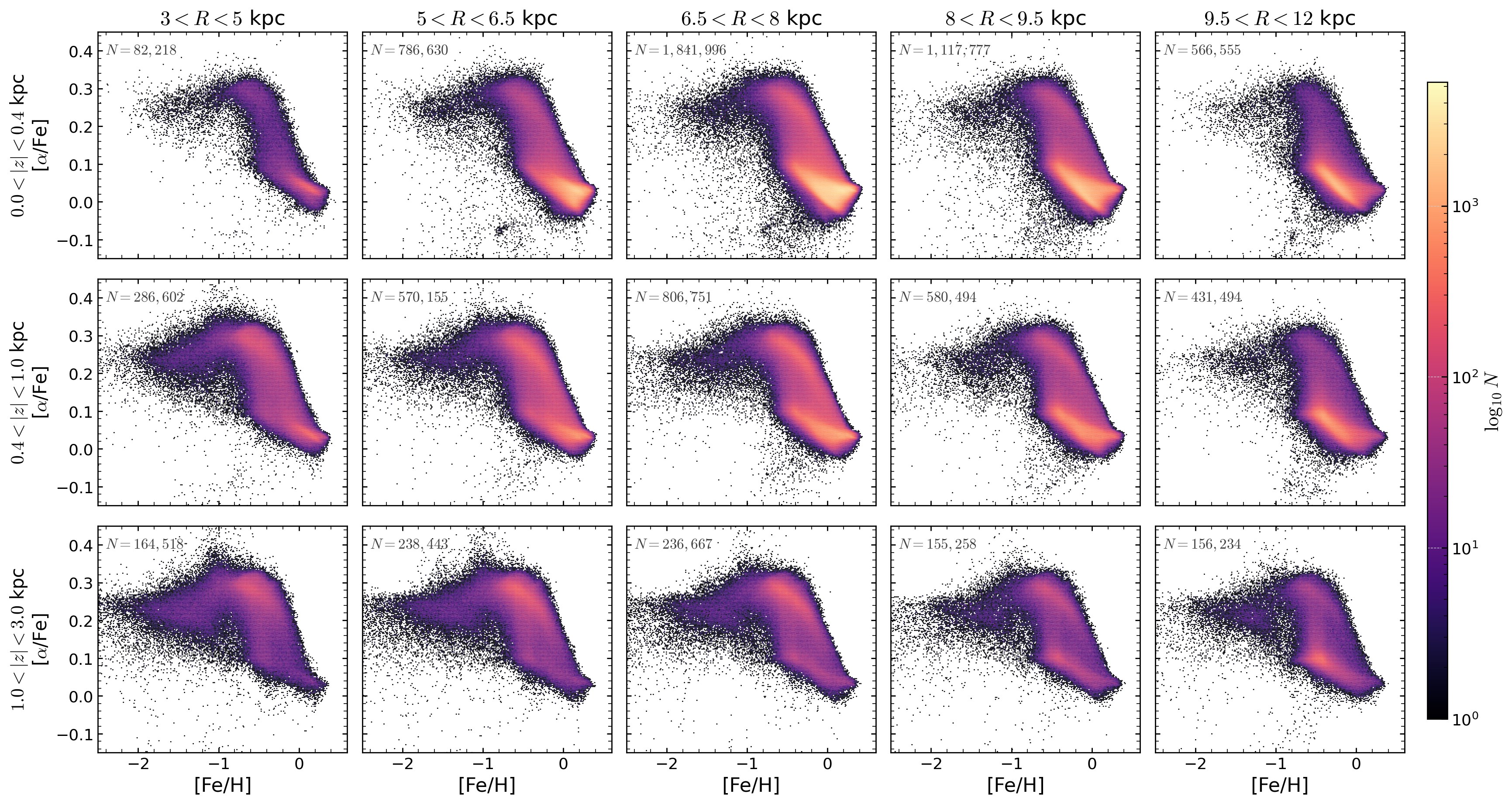}
\caption{The [$\alpha$/Fe]--[Fe/H] plane of $8.0$ million Gaia XP red giant stars ($T_{\rm eff}<5300$\,K, $\log g<3.5$) with median per-pixel S/N $\ge 64$ and full phase-space information, in bins of Galactocentric radius $R$ in the midplane and height $|z|$. The elemental abundances are predicted from the aligned XP embedding of the JEPA model. The main chemical structures of the MW are recovered from resolving power $\sim50$ XP spectra alone. \label{fig:xpplane}}
\end{figure*}

\section{Aligned embedding gives a physical coordinate system}
\label{sec:coord} 

With the proper alignment the leading variance is a good indicator of the physics. Figure~\ref{fig:coord} colors the leading principal-component plane of the 4-survey JEPA embedding by surface gravity, [$\alpha$/Fe], and chemical category, which together identify what each direction encodes. Surface gravity varies smoothly across that plane (left), separating the dwarf and giant branches, which confirms that the Kiel-shaped plane of Figure~\ref{fig:obj} is gravity-ordered. The residual deviation from the true Kiel diagram is a mild nonlinear warp on top of a dominant linear transformation.

The chemically defined populations (right) then organize along the iron-abundance direction of the arc of giant stars. Position along that arc tracks [Fe/H], and the iron-abundance axis of the embedding places the populations exactly in the same order as does their chemistry, with the median third-component value falling from $+5.7$ for the thin disk, through the thick disk and the low-$\alpha$\,\&\,metal-poor stars, to $-22.8$ for the accreted stars, as their median [Fe/H] falls from $-0.20$ to $-1.29$. 

How separable a population is depends on where it falls in this plane. The two disks interleave as curved bands. A nonlinear nearest-neighbor classifier using only their positions in the plane separates them with and an area under the receiver operating characteristic curve (AUC) of $0.83$, where $0.5$ is chance and 1 is perfect separation, while a linear logistic regression reaches only 0.55.
The accreted stars, by contrast, separate from the thick disk at AUC $=0.97$. 

The [$\alpha$/Fe] coloring (middle) alternates along this same sequence (median $+0.05$, $+0.27$, $-0.01$, $+0.19$ across the four populations), and the middle panel therefore shows interleaved [$\alpha$/Fe] bands rather than a single smooth gradient. The iron-abundance axis does saturate in the sparse very-metal-poor tail, where the third component changes little between [Fe/H] $\approx$ $-1.3$ and $-2.4$, but the very metal-poor stars remain visible, clustered at the extreme tip of the arc.

These [$\alpha$/Fe] bands are real but folded, since a nonlinear read-out of Kiel-shaped plane position recovers $R^2=0.32$ of [$\alpha$/Fe], whereas a linear read-out of the same two coordinates recovers none of it ($R^2 = 0.01$). The [$\alpha$/Fe] content from the plane alone is a compressed trace of population-level correlations, and the alignment adds a linear layout along the higher principal-component axes. With all training labels, a ridge fit recovers $R^2=0.26$ of [$\alpha$/Fe] from the first three principal components, $0.67$ from the first 30, and 0.80 from all 768.

The leading components therefore come in a definite order, with the Kiel plane first, iron abundance next, finer chemistry deeper. The alignment has thus spent its leading variance along the known stellar-parameter and chemical-evolution axes, which is the ordering the reconstructive AE baseline does not produce.

Due to the fact that the physics lies on the readable axes, abundances can be predicted for stars far beyond the labeled cross-match. We put all public $34.5$ million Gaia DR3 XP sampled spectra through a four-survey JEPA backbone, a variant retrained with regularizer weight $\lambda=10$ whose construction and validation will be presented in a future paper (Zhao et al. in prep), where we also provide [Fe/H] and [$\alpha$/Fe] from the XP embedding with per-label nonlinear probes, the same as in Section~\ref{sec:methods}. 

Selecting red giants ($T_{\rm eff}<5300$\,K, $\log g<3.5$) yields $13.0$ million stars, of which $8.7$ million have median per-pixel S/N of the sampled spectrum $\ge 64$. Figure~\ref{fig:xpplane} shows the $8.0$ million of these with full phase-space information, using the photo-geometric distances of \citet{2021AJ....161..147B}, within $3<R<12$\,kpc and $|z|<3$\,kpc. The locus of the density in the Tinsley--Wallerstein plane recovers the established chemical structure of the Milky Way from resolving power $\sim 50$ spectra alone, including the low-$\alpha$ and high-$\alpha$ sequences, and Gaia--Enceladus/Sausage. This shows that Gaia XP alone can already support disk and halo studies.

\section{Discussion} \label{sec:disc}

Multi-view identifiability theory provides a partial explanation for these findings \citep{vonKugelgen2021,Daunhawer2023}. Different survey views of the same star share an underlying physical state, so an objective that encourages agreement between matched views, while preventing collapse, can preserve shared physical information and suppress instrument-specific information. However, the theory does not predict that the recovered information is linearly readable or lies on the leading principal axes. Moreover, these results for networks like ours assume a contrastive objective rather than the non-contrastive regularization used here. For JEPA specifically, \citet{Klindt2026} recently proved linear identifiability in a single-instrument setting, whereas our multi-instrument experiments lie outside that setting. More theoretical work is needed to explain its success.

Stellar spectra are a favorable place to test the validity or limitation of such theories. The latent variables that generate a spectrum are few, known, and physically meaningful, and each is measured independently at higher resolution, so a representation can be checked against the physics. Identifiability theory is otherwise mainly tested on simulations, where the latent state is whatever the model says it is. Several spectrographs observing the same star instead supply independent instrumental views of a state that is real and externally known, which is what makes the recovered axes interpretable, and what makes the field of stellar spectroscopy a proving ground for these methods rather than only a consumer of them.

Reconstruction and contrastive training have been two common approaches in astronomy, serving jointly or separately depending on the needs. A training objective aligned with the physical structure of the data rather than with the pixels, as well as requiring no negative pairs, has been identified as one of the clearer paths for progress \citep{2026arXiv260802573Z}. Cross-survey agreement using JEPA is such a task. This carries practical consequences for Galactic archaeology, namely a specific prescription for building cross-survey models. As shown above, alignment without negatives scales naturally as surveys are added, removes the need for large batches constructed to supply negative examples, and, in our experiments, produces an embedding whose leading axes can be read directly rather than only through a further trained network. In a complementary approach, \citet{2026arXiv261007112T} built label-free coordinates for APOGEE and DESI spectra using manifold learning applied separately to each survey, without neural networks. They found the APOGEE coordinates to be close to linear in $T_{\rm eff}$ and $\log g$. Here, the linear legibility of stellar parameters instead emerges from aligning learned embeddings across surveys.

Several conditions bound these findings. First, all read-outs are relative to a cross-matched, magnitude- and color-selected sample, so the category fractions reflect the sample's selection function. Second, the embedding is from extinction-sensitive Gaia XP, and extinction is shared by all views of a given star. It can therefore remain in the representation alongside the stellar parameters and we have not explicitly removed it. Third, our alignment starts from encoders pre-trained separately on larger datasets within a specific survey. This is sufficient for our purpose, which is to ask how alignment reorganizes the information already present in valid encoders. An interesting investigation would be to train the encoders and alignment networks jointly using only the cross-matched subsets. Fourth, our analysis uses the embeddings from the survey-specific head introduced in Section~\ref{sec:methods}, which we found to perform better than the projected embeddings on which the loss terms act. The influence of the projector architecture on the preceding survey-specific embeddings has not been systematically investigated. Lastly, we study only the explicit alignment. Whether implicit alignment through a shared tokenizer (e.g. \citealp{Shen2025}), or an omnimodal model mediated by discrete tokens (AION; \citealp{Parker2025}), is equally linearly legible remains open. 

\section{Summary}
\label{sec:summary}

In this paper, we ask what physical structure can be gained from multi-survey alignment and how strongly that structure depends on the alignment objective. We begin with reconstructive autoencoders that produce the first-stage survey embeddings, and compare two ways of aligning them in a shared space, particularly contrastive alignment with CLIP and negative-free alignment based on the core ideas of JEPA. Our central question is not only whether alignment improves downstream prediction, but whether it reorganizes the representation so that known stellar and Galactic structure becomes directly legible.

We find that JEPA reorganizes the information contained in the autoencoder embeddings particularly effectively: 
\begin{itemize}

    \item Alignment across multiple spectroscopic surveys brings effective temperature and surface gravity into the first two principal components. Their organization in the PC1-PC2 plane forms a close analog of the Kiel diagram (Figure~\ref{fig:obj}), despite no stellar labels being used during alignment. Iron abundance is strongly associated with the third principal component.
    
    \item Linear probes recover these stellar parameters accurately from aligned representation, including in label-scarce settings and when restricted to only the first three principal components.
    
    \item The negative-free objective extends naturally as additional surveys are included and improves the resulting representation without requiring the large batches of negative pairs used by CLIP.
    
    \item The physical interpretation of the Kiel-like plane analog is reinforced by its gravity ordering. Coloring the representation by iron abundance and [$\alpha$/Fe], and applying corresponding population divisions, separates familiar Galactic populations including the thin and thick disk, Gaia--Enceladus/Sausage, and very metal-poor tail (Figure~\ref{fig:coord}). [$\alpha$/Fe] is likewise linearly accessible from the embeddings. Together, these results show that the aligned embedding forms a physically organized coordinate system in which Galactic chemical evolution become legible.
    
    \item As a direct test beyond the training and withheld test sample, applying the representation to the public full-sky Gaia DR3 XP data recovers the corresponding Galactic components as functions of Galactic radius and height (Figure~\ref{fig:xpplane}).
    
\end{itemize}
CLIP and JEPA both improve upon the single-survey autoencoder, while JEPA is consistently superior to CLIP in the tests presented here, for Gaia XP. The alignment objective, rather than alignment alone, therefore determines whether the physical structure encoded in Gaia XP becomes accessible to a simple linear read-out. These results suggest that the negative-free alignment can provide an interpretable, scientifically useful, and scalable approach to representation learning in astronomy. 

During the preparation of this paper, related ideas were independently demonstrated within individual modalities for radio galaxies and light curves \citep{2026arXiv260830594L, 2026arXiv260628446R}, providing astronomical analogs of LeJEPA \citep{Balestriero2025}. Here we show that the same principle extends naturally to the multi-survey setting. Since matched observations of the same object already provide distinct physical views, no artificial cropping or augmentation is required to construct the correspondence. Multi-survey spectroscopy therefore provides a particularly direct example.

The broader implication concerns how such structure can be discovered. These models were given no information about the Kiel diagram, iron abundance, or the thin and thick disks. They were given only the fact that particular spectra correspond to the same star. Recovering those structures along the leading axes therefore acts as a validation, since the representation rediscovers physics established independently. The same procedure can in principle be used in the opposite direction. When the governing structure is not already known, the combination of several instruments observing the same objects, an objective that preserves what those views share, and inspection of the resulting leading axes, together provide a way for structure to emerge, without prescribing in advance what should be found. Nothing in this procedure is specific to stars or spectra. It requires only that the same object be observed in more than one way, a setting increasingly common across observational science.

The space of label-free objectives remains large and largely unexplored. Our results show that the choice of alignment objective is a first-order design decision when building a cross-survey model, and motivate the design of objectives for interpretable representations rather than for benchmark accuracy alone.

\section*{Acknowledgements}
X.Z., R.F.G.W., and A.S.Sz. have been supported by a grant from Schmidt Sciences. Y.S.T. is supported by the National Science Foundation under Grant No. AST-2406729 and by a Humboldt Research Award from the Alexander von Humboldt Foundation. We acknowledge the use of Anthropic's Claude models and OpenAI's ChatGPT models for code development, data-analysis assistance, and manuscript editing. All AI-assisted content was reviewed, verified, and further edited by the authors, who take full responsibility for the accuracy and integrity of this work.

This work made use of the data from LAMOST (Large Sky Area Multi-Object Fiber Spectroscopic Telescope, also known as the Guoshoujing Telescope) (https://cstr.cn/31118.02.LAMOST). LAMOST is a Chinese national mega-science facility, operated by National Astronomical Observatories, Chinese Academy of Sciences.

This research used data obtained with the Dark Energy Spectroscopic Instrument (DESI). DESI construction and operations is managed by the Lawrence Berkeley National Laboratory. This material is based upon work supported by the U.S. Department of Energy, Office of Science, Office of High-Energy Physics, under Contract No. DE–AC02–05CH11231, and by the National Energy Research Scientific Computing Center, a DOE Office of Science User Facility under the same contract. Additional support for DESI was provided by the U.S. National Science Foundation (NSF), Division of Astronomical Sciences under Contract No. AST-0950945 to the NSF’s National Optical-Infrared Astronomy Research Laboratory; the Science and Technology Facilities Council of the United Kingdom; the Gordon and Betty Moore Foundation; the Heising-Simons Foundation; the French Alternative Energies and Atomic Energy Commission (CEA); the National Council of Humanities, Science and Technology of Mexico (CONAHCYT); the Ministry of Science and Innovation of Spain (MICINN), and by the DESI Member Institutions: www.desi.lbl.gov/collaborating-institutions. The DESI collaboration is honored to be permitted to conduct scientific research on I’oligam Du’ag (Kitt Peak), a mountain with particular significance to the Tohono O’odham Nation. Any opinions, findings, and conclusions or recommendations expressed in this material are those of the author(s) and do not necessarily reflect the views of the U.S. National Science Foundation, the U.S. Department of Energy, or any of the listed funding agencies.

Funding for the Sloan Digital Sky Survey V has been provided by the Alfred P. Sloan Foundation, the Heising-Simons Foundation, the National Science Foundation, and the Participating Institutions. SDSS acknowledges support and resources from the Center for High-Performance Computing at the University of Utah. SDSS telescopes are located at Apache Point Observatory, funded by the Astrophysical Research Consortium and operated by New Mexico State University, and at Las Campanas Observatory, operated by the Carnegie Institution for Science. The SDSS web site is www.sdss.org. 

SDSS is managed by the Astrophysical Research Consortium for the Participating Institutions of the SDSS Collaboration, including the Carnegie Institution for Science, Chilean National Time Allocation Committee (CNTAC) ratified researchers, Caltech, the Gotham Participation Group, Harvard University, Heidelberg University, The Flatiron Institute,  The Johns Hopkins University, L’Ecole polytechnique fédérale de Lausanne (EPFL), Leibniz-Institut für Astrophysik Potsdam (AIP), Max-Planck-Institut für Astronomie (MPIA Heidelberg), Max-Planck-Institut für Extraterrestrische Physik (MPE), Nanjing University, National Astronomical Observatories of China (NAOC), New Mexico State University, The Ohio State University, Pennsylvania State University, Smithsonian Astrophysical Observatory, Space Telescope Science Institute (STScI), the Stellar Astrophysics Participation Group, Universidad Nacional Autónoma de México (UNAM), University of Arizona, University of Colorado Boulder, University of Illinois at Urbana-Champaign, University of Toronto, University of Utah, University of Virginia, Yale University, and Yunnan University.

\section*{Data Availability}
All observational data used in this work are publicly available from the archives. No proprietary data were used. 
The code builds on SpecCLIP (\url{https://github.com/Xiaosheng-Zhao/SpecCLIP}), whose Gaia XP and LAMOST LRS encoders and their weights are public; the full code and trained weights of this work will be released upon acceptance of the paper.
  
\software{
    PyTorch \citep{NEURIPS2019_9015},
    Astropy \citep{2013A&A...558A..33A, 2018AJ....156..123A},
    scikit-learn \citep{JMLR:v12:pedregosa11a},
    galpy \citep{2015ApJS..216...29B},
    SpecCLIP \citep{Zhao2025,zhao_2025_17824840}
}

\appendix
\restartappendixnumbering

\section{Training objectives and variants}
\label{sec:losses} 

This appendix records the training objectives that we tested. We do not attempt a comprehensive survey of JEPA variants. Rather, we compare the non-contrastive joint-embedding objectives that were most relevant to our experiments, together with CLIP and an autoencoder baseline. The main text defines the objective adopted in our analysis. CLIP uses the standard contrastive loss \citep{Oord2018, Radford2021},
and AE is the Gaia-XP autoencoder with no cross-survey alignment term. 

The non-contrastive objectives share the invariance term of Equation~\ref{eq:inv}, applied to the projected embedding $\mathbf{z}=g(\mathbf{h})$ of Section~\ref{sec:objective}, and differ primarily in the regularizer that prevents the trivial solution $\mathbf{z}\equiv\mathrm{const}$. We tested the following three regularizers: 

\begin{itemize}
    \item RDMReg (Rectified $L_p$JEPA) matches the projected embedding distribution to a rectified generalized Gaussian target
    using a sliced 2-Wasserstein distance. In the adopted configuration
    the target is a rectified Laplace ($p=1$), which favors sparse, non-negative coordinates, with a coordinate being exactly zero for a substantial fraction of stars and positive otherwise.
    \item SIGReg (LeJEPA) matches the projected embedding distribution to an isotropic Gaussian target using the Epps-Pulley statistic \citep{Balestriero2025}, which compares empirical and target characteristic functions. It therefore favors a dense, isotropic embedding distribution rather than sparse non-negative coordinates.
    \item VICReg \citep{Bardes2021} does not prescribe a full target distribution. Its variance term keeps each coordinate from collapsing below a prescribed scale, while its covariance term suppresses correlations between different coordinates. It therefore constrains second-order statistics rather than the full projected distribution. 
\end{itemize}

All three objectives align the surveys in our experiments, but SIGReg and VICReg leave the chemical structure less concentrated in the leading-embedding (of the survey-specific head) variance and have slightly worse probe performance than RDMReg. We therefore adopt RDMReg throughout the main analysis, and its exact four-survey form and weights are given in Appendix~\ref{sec:app}.

\section{Implementation} \label{sec:app}

The two-survey models use the two SpecCLIP frozen backbones, the Gaia-XP overcomplete autoencoder and the LAMOST-LRS masked-transformer encoder \citep{Zhao2025}. The XP backbone takes 343-point XP spectra, each divided by its flux at 550~nm, as inputs and returns a single 768-d latent vector, which a trainable MLP head maps into the shared 768-d space. We keep the encoder bottleneck overcomplete (768 dimensions for a 343-point input). \citet{Zhao2025} found embeddings of 256 to 768 dimensions to perform comparably, with those above the input size marginally better (their Table~3), while stronger compression discards chemical information. Retraining the same autoencoder with 24-, 12- and 6-dimensional bottlenecks lowers the [$\alpha$/Fe] $R^2$ of the nonlinear probe from $0.75$ to $0.57$, $0.47$ and $0.36$ ($N=10{,}000$). The LAMOST-LRS backbone takes 1{,}462-pixel continuum-normalized spectra as inputs and returns a sequence of 146 tokens of width 768, which a trainable cross-attention head with one learned query pools into the same space. The four-survey models add two more frozen backbones and heads. The first is a DESI masked-transformer encoder with the same architecture and head as LAMOST LRS; DESI spectra are continuum-normalized in the same way and interpolated onto the LAMOST-LRS wavelength grid.

\begin{figure*}
\centering
\includegraphics[width=\textwidth]{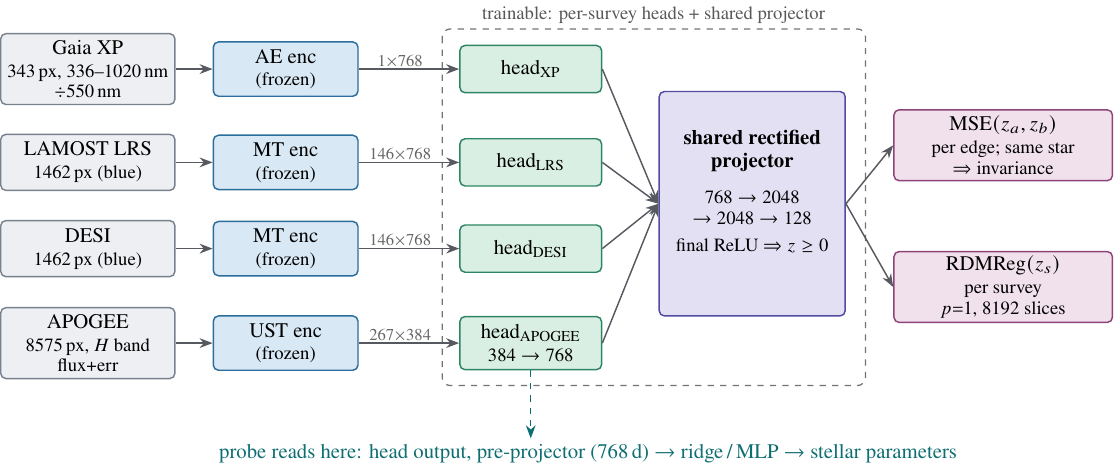}
\caption{Architecture of four-survey JEPA.  Each survey spectrum is encoded by a frozen reconstructive backbone (an overcomplete autoencoder for Gaia XP, masked transformers for LAMOST LRS and DESI, and a wavelength-aware autoencoder built on the universal spectral tokenizer for APOGEE; labeled AE, MT and UST, respectively) and is then mapped onto a common 768-dimensional representation by a trainable head: an MLP for the single XP latent vector, and cross-attention pooling for the token sequences of the other three surveys. A single rectified projector, shared by all four surveys, maps these representations into a 128-dimensional non-negative space. A mean-squared invariance term aligns the representations across the six cross-matched survey edges, pulling different views of the same star towards a common point. A per-survey rectified distribution-matching regularization (RDMReg) prevents collapse constraint without requiring negative pairs.  Only the heads and the shared projector are trained (33.8~M parameters, against 142~M held frozen).  All downstream probes read the head output before the projector (dashed arrow), so the recovered stellar parameters come from the 768-dimensional head representation rather than from the loss-space projection. \label{fig:arch_details}}
\end{figure*}

The second is an APOGEE wavelength-aware autoencoder, a $384$-d transformer of depth $8$ with $8$ attention heads that reads the spectrum on its native wavelength grid by attaching continuous Fourier wavelength embeddings to each pixel. The per-pixel flux uncertainty enters as a second input channel, and the local line-spread-function width (the FWHM $\lambda/R$ from a per-survey resolution model $R(\lambda)$ with $R = 22{,}500$, constant for APOGEE) is encoded by a further Fourier embedding added to each token, so that resolution is given rather than inferred. The $8{,}575$-d continuum-normalized spectrum is patchified into $267$ tokens of $32$ pixels each (the trailing 31 pixels are dropped) and trained by reconstruction alone \citep[the universal-spectral-tokenizer idea of][]{Shen2025}, its frozen tokens then read into the shared 768-d space by a trainable cross-attention head. 
Every objective uses this architecture, so across the comparison only the loss differs. Figure~\ref{fig:arch_details} shows the four-survey architecture.

\paragraph{Four-survey JEPA objective.} For each survey $s\in\mathcal{S}=\{\mathrm{Gaia\,XP},\mathrm{LAMOST\,LRS},\mathrm{DESI},\mathrm{APOGEE}\}$, the frozen backbone encoder $E_s$ and trainable head $H_s$ produce $\mathbf{h}_s=H_s(E_s(\mathbf{x}_s))\in\mathbb{R}^{768}$. A shared projector $g$, with layers $768\to2048\to2048\to128$, batch normalization and a ReLU after each hidden layer, and a final ReLU, produces $\mathbf{z}_s=g(\mathbf{h}_s)\in\mathbb{R}^{128}_{\ge0}$. Alignment runs over all six survey pairs, referred to as edges, $e=(a_e,b_e)\in\mathcal{E}$. Each training step draws $B=128$ matched pairs from each edge. Since the six matched-pair datasets differ in size, an epoch is one pass through the largest dataset, XP--LRS, reusing samples from smaller datasets as needed. The invariance loss is the mean squared difference per coordinate, averaged across edges with weights $w_e$,
\begin{equation}
\mathcal{L}_{\rm inv}=\frac{\sum_{e}w_e\,(Bd)^{-1}\sum_{i=1}^{B}\lVert\mathbf{z}_{a_e}^{(i)}-\mathbf{z}_{b_e}^{(i)}\rVert^2}{\sum_{e}w_e},
\end{equation}
with $d=128$, $w_e=2$ for the XP--DESI and XP--APOGEE edges, and $w_e=1$ otherwise. The regularizer is applied separately to each survey. Since each survey appears in three edges, each step yields $N_s=3B$ projected embeddings of survey $s$. These embeddings are compared with $N_s$ independently drawn target vectors $\mathbf{t}$, whose coordinates are $t_c=\max(0,\epsilon_c)$, where $\epsilon_c$ follows a zero-mean, unit-variance Laplace distribution:
\begin{equation}
\mathcal{R}_s=\frac{1}{K}\sum_{k=1}^{K}\frac{1}{N_s}\sum_{j=1}^{N_s}\Bigl[(\mathbf{u}_k^{\!\top}\mathbf{z}_s)_{(j)}-(\mathbf{u}_k^{\!\top}\mathbf{t})_{(j)}\Bigr]^2,
\end{equation}
where $(\cdot)_{(j)}$ denotes the $j$-th smallest value among the $N_s$ projections, and the $K=8192$ unit vectors $\mathbf{u}_k$ are sampled uniformly from the unit sphere, independently for each survey at each step. The total loss is $\mathcal{L}=25\,\mathcal{L}_{\rm inv}+125\,|\mathcal{S}|^{-1}\sum_{s}\mathcal{R}_s$, corresponding to a regularization-to-invariance weight ratio of $\lambda=5$. The two-survey JEPA model uses the same objective with a single edge and $B=256$. We train the four-survey models with AdamW \citep{2017arXiv171105101L}, with learning rate $10^{-4}$ and weight decay $10^{-2}$, a cosine learning-rate schedule, bf16 mixed precision, and gradient-norm clipping at $1$. The two-survey models keep the original SpecCLIP training configuration, which differs only in weight decay ($0.05$), AdamW $\beta_2=0.95$, and a short linear warm-up.

\paragraph{Four-survey CLIP objective.} The contrastive control uses the same backbones, trainable heads, survey edges and edge weights as JEPA. It also uses the same projector, but omits the final ReLU and normalizes its outputs to unit length, $\hat{\mathbf{z}}=\mathbf{z}/\lVert\mathbf{z}\rVert$. For each edge $e$ and a batch of $B$ matched pairs, the similarity matrix is defined as $S^{(e)}_{ij}=\hat{\mathbf{z}}_{a_e}^{(i)\top}\hat{\mathbf{z}}_{b_e}^{(j)}/\tau$. A symmetric cross-entropy loss identifies the matched star among the $B$ candidates in each direction,
\begin{equation}
\ell_e=-\frac{1}{2B}\sum_{i=1}^{B}\left[\ln\frac{e^{S^{(e)}_{ii}}}{\sum_{j}e^{S^{(e)}_{ij}}}+\ln\frac{e^{S^{(e)}_{ii}}}{\sum_{j}e^{S^{(e)}_{ji}}}\right],
\qquad
\mathcal{L}_{\rm CLIP}=\frac{\sum_e w_e\,\ell_e}{\sum_e w_e}.
\end{equation}
The temperature $\tau$ is a single learned parameter shared across all edges, initialized at $0.07$ and bounded below by $0.01$. The four-survey CLIP model uses $B=384$ matched pairs per edge, compared with $B=1024$ for the two-survey experiment. All other optimization settings match those of JEPA.

Checkpoint selection follows the same rule for every objective, which is trained with the same schedule, and training continuously tracks the best-validation checkpoint alongside periodic snapshots. The five two-survey models are selected at the epoch of lowest label-free validation loss on the withheld stars split from the training set, a criterion that never sees a reference label. 

For the four-survey JEPA the label-free validation loss is not a usable selector, because it is lowest at epoch~$0$, before alignment reshapes the embedding, and drifts upward as the distribution-matching regularizer acts. The drift is a property of the loss rather than of the model because the invariance term is an unnormalized squared distance, so its magnitude grows as the regularizer inflates the embedding from its small, near-collapsed initial state toward the fixed-width target. The early epochs therefore show a deceptively low loss that measures the embedding's scale as much as its quality, even while probe recovery improves. The model is therefore chosen by probe accuracy on the reference labels. 

This asymmetry could favor the four-survey model, but it does not drive the central result. The two-survey JEPA, selected purely by the label-free validation loss with no reference to the labels, already leads the two-survey CLIP counterpart and the autoencoder on the discriminating linear [$\alpha$/Fe] axis (Table~\ref{tab:r2}), so the performance ordering is independent of the selection criterion. The four-survey CLIP control is selected the same way as the four-survey JEPA, by probe accuracy, so the comparison of Section~\ref{sec:linear} treats the two four-survey models symmetrically. One training asymmetry favors the contrastive runs. Specifically, the contrastive runs use larger batches (1024 against 256 at two surveys, and 384 against 128 at four), which benefits a loss whose performance depends on the number of available negatives.

The ridge penalty strength is chosen from $\{0.1,1,10,10^2,10^3\}$ by leave-one-out cross-validation among the labeled training set. The MLP has layer widths 768, 256, 128, and 1, with ReLU \citep{3104322.3104425} activations and dropout $0.1$, and is trained with Adam \citep{Kingma2014AdamAM} at learning rate $10^{-3}$ and weight decay $10^{-4}$ for a fixed $200$ epochs with no early stopping. The learned contrastive temperature converges to $\tau=0.042$ in the two-survey and $\tau=0.028$ in the four-survey run.

\bibliographystyle{aasjournalv7}
\bibliography{refs}

\end{document}